\documentclass{article}
\usepackage{dcase2024,amsmath,graphicx,url,times,booktabs, tabularx, amssymb, bm}
\usepackage{url}
\usepackage{lipsum}  
\usepackage{tipa}
\usepackage[inline]{enumitem}
\usepackage{xkeyval}
\usepackage{calc}
\usepackage{subcaption}
\usepackage{cite}

\title{Angular Distance Distribution Loss for Audio Classification}

\name{
   Antonio Almudévar$^{1}\sthanks{This work was supported by MCIN/AEI/10.13039/501100011033 under Grants PDC2021-120846-C41 \& PID2021-126061OB-C44, and in part by the Government of Aragón (Grant Group T36 23R). This project has received funding from the European Union’s Horizon 2020 research and innovation programme under the Marie Skłodowska-Curie grant agreement No 101007666.}$,
   Romain Serizel$^{2}$, 
   Alfonso Ortega$^{1}$
}
\address{
    $^1$ ViVoLab, Aragón Institute for Engineering Research (I3A), University of Zaragoza, Spain\\       
    $^2$ Université de Lorraine, CNRS, Inria, LORIA, F-54000 Nancy, France\\
    almudevar@unizar.es
}

\begin{document}

\ninept
\maketitle

\begin{sloppy}

\begin{abstract}
    Classification is a pivotal task in deep learning not only because of its intrinsic importance, but also for providing embeddings with desirable properties in other tasks. To optimize these properties, a wide variety of loss functions have been proposed that attempt to minimize the intra-class distance and maximize the inter-class distance in the embeddings space. In this paper we argue that, in addition to these two, eliminating hierarchies within and among classes are two other desirable properties for classification embeddings. Furthermore, we propose the Angular Distance Distribution (ADD) Loss, which aims to enhance the four previous properties jointly. For this purpose, it imposes conditions on the first and second order statistical moments of the angular distance between embeddings. Finally, we perform experiments showing that our loss function improves all four properties and, consequently, performs better than other loss functions in audio classification tasks.
\end{abstract}


\section{Introduction} \label{sec:intro}
Classification is one of the main tasks to be solved with machine learning. In this task, there are typically high-dimensional elements and the goal is to decide to which class of a finite set each of these elements belongs. For this purpose, most of the solutions, particularly those based on deep learning, involve obtaining intermediate representations of reduced dimension of the elements to be classified. These representations are called embeddings and they can be considered as a summary of these elements containing the information that is relevant for classification. This problem is very popular not only because of its intrinsic importance, but also because it provides a simple way to obtain embeddings compared to other methods. Embeddings are useful for a multitude of tasks such as anomaly detection \cite{chandola2009anomaly, kawaguchi2021description}, biometric recognition \cite{deng2019arcface, wang2018cosface}, etc. The standard loss function to solve the classification task is the cross-entropy. As a secondary result of using this loss function, the embeddings of the different classes usually end up being somewhat separated. However, it is common to impose certain conditions directly on them due to two reasons:
\begin{enumerate*}[label=(\roman*)]
  \item this tends to improve the performance in the classification problem by guiding more the optimization \cite{ranasinghe2021orthogonal, almudevar23_interspeech}; and
  \item it may be desirable for embeddings to have certain properties when used for a specific task other than classification \cite{lin2017focal, LiuCQUPT2022}.
\end{enumerate*} 
These conditions on embeddings are usually imposed through the loss function. Typically, a term is added to the cross-entropy or a modification is made to it.

In this paper we propose a loss function that is added to cross-entropy and we call it Angular Distance Distribution Loss because it imposes conditions on the first and second order statistical moments of the angular distances between embeddings in order to organize the embeddings in the space. Specifically, this organization consists of:
\begin{enumerate*}[label=(\roman*)]
  \item bringing embeddings of the same class closer,
  \item moving embeddings of different class away,
  \item minimizing the variation of the distances of the embeddings of the same class, and
  \item making the embeddings of a class equal in distance to the embeddings of any class.
\end{enumerate*} 
Traditionally, only the first two have been considered in the literature. However, in section \ref{sec:proposed_method} we formalize all four, arguing why they are all important. In addition, we reason how they relate to the statistical moments of the distances between embeddings.
Furthermore, we propose an experimental framework with different Audio Classification datasets. In these experiments, on the one hand, we verify that our embeddings satisfy the properties described in the previous paragraph, so we verify that our loss function encourages the properties to be satisfied. On the other hand, we obtain a better accuracy than other loss functions that aim to establish conditions on the embeddings. Thus, we verify that the described properties translate into better classification performance.
The details of these experiments are presented in the section \ref{sec:experiments} and can be replicated using the code in
\url{https://github.com/antonioalmudevar/distance_distribution_loss}

\section{Related Work} \label{sec:related_work}
\textbf{Audio Classification} consists of identifying to which class an audio belongs \cite{gemmeke2017audio, mesaros2017dcase}. In recent years it has received a lot of interest from the community \cite{turpault2019sound, ronchini2021impact}. Solutions to this problem typically involve an embedding extractor followed by a small classifier net which are trained by minimizing cross-entropy. In many SOTA solutions the embedding extractor has a large number of parameters, so it is common to pre-train it with a large dataset and perform finetuning for the desired dataset. Although convolutional architectures has been widespread used \cite{hershey2017cnn, cakir2017convolutional, serizel2018large}, the most popular systems nowadays are transformer-based. These include Audio Spectrogram Transformer (AST) \cite{gong2021ast} and BEATs \cite{chen2022beats}.

\vspace{1mm}\noindent\textbf{Loss Functions.} It has been observed in different works that separating the embeddings of different classes often results in better performance in the classification task \cite{sun2020fixing, zhang2020rbf, ranasinghe2021orthogonal, sheth2024auxiliary}. Two loss functions that stand out are Focal Loss \cite{lin2017focal} and Orthogonal Projection Loss (OPL) \cite{ranasinghe2021orthogonal}, with which we compare our proposal.


\section{Proposed Method} \label{sec:proposed_method}
The problem we seek to solve in this paper is that of canonical classification, which has two characteristics:
\begin{enumerate*}[label=(\roman*)]
  \item all errors are considered equally critical; and
  \item all elements are considered equally similar to each other within a class.
\end{enumerate*}
This means that intra-class and inter-class hierarchies are not desirable. In fact, the standard evaluation metric is accuracy, which considers all errors and correct predictions equally relevant. The presence of these hierarchies is desirable in some scenarios, but our goal is not to solve the latter.

\vspace{-5pt}
\subsection{Classification Solution Formulation}
Let $\mathcal{D}=\{x^{(i)},y^{(i)}\}_{i=1}^N$ be the dataset, where $x^{(i)}$ is each input and $y^{(i)}$ the label of $x^{(i)}$ and is a vector containing at each position $j$ the probability that $x^{(i)}$ belongs to class $j$. The objective is to design a system that allows us to obtain a prediction of $y^{(i)}$ which we denote $\tilde{y}^{(i)}$.
Typical deep learning classifier solutions consists of
\begin{enumerate*}[label=(\roman*)]
  \item an embeddings extractor $f_\theta$, which provides the embedding as $z^{(i)} = f_\theta(x^{(i)}) \in \mathbb{R}^k$; and
  \item a classifier net $g_\phi$, which gives the predictions as $\tilde{y}^{(i)}=g_\phi(z^{(i)}) \in \mathbb{R}^{c}$. 
\end{enumerate*}
Cross-entropy between $y_i$ and $\tilde{y}_i$ is used as loss function, which we call $\mathcal{L}_{CE}$.

\vspace{-5pt}
\subsection{Desirable Properties of Embeddings} \label{subsec:properties}
We explain below some desirable properties of embeddings for classification. In figs. \ref{fig:intra_clust} to \ref{fig:inter_equi} the dots correspond to low dimensional representations of the embeddings and different colors are used to indicate different classes.
\begin{itemize}
    \item \textbf{Intra-class clustering}: The embeddings of the same class are close to each other in space. This has been shown to improve performance in classification and additional tasks.
        \vspace{-3mm}
        \begin{figure}[ht!]
          \centering
          \begin{subfigure}[b]{0.3\columnwidth}
            \includegraphics[width=\textwidth]{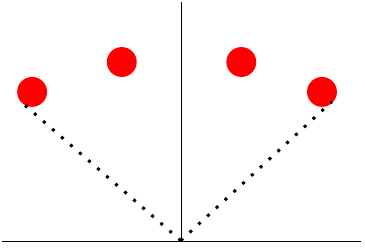}
          \end{subfigure}
          \hspace{3mm}
          \begin{subfigure}[b]{0.3\columnwidth}
            \includegraphics[width=\textwidth]{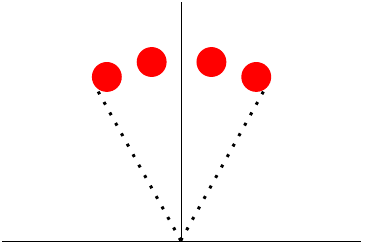}
          \end{subfigure}
          \vspace{-2mm}
          \caption{Low (left) and high (right) Intra-class clustering}
          \label{fig:intra_clust}
        \end{figure}
        \vspace{-3mm}
        
    \item \textbf{Intra-class equidistance}: All the embeddings from the same class have approximately the same distance from each other. From a conceptual perspective, all the elements in a given class should be equally similar.
        \vspace{-3mm}
        \begin{figure}[ht!]
          \centering
          \begin{subfigure}[b]{0.3\columnwidth}
            \includegraphics[width=\textwidth]{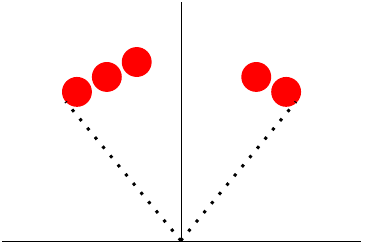}
          \end{subfigure}
          \hspace{3mm}
          \begin{subfigure}[b]{0.3\columnwidth}
            \includegraphics[width=\textwidth]{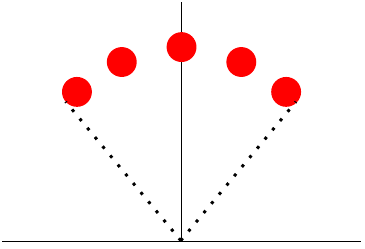}
          \end{subfigure}
          \vspace{-2mm}
          \caption{Low (left) and high (right) Intra-class equidistance}
          \label{fig:intra_equi}
        \end{figure}
        \vspace{-3mm}
        
    \item \textbf{Inter-class separation}: Embeddings of different classes are far away from each other. This allows to take better advantage of all the space and, as a consequence, improves the performance in different tasks, especially when coupled with intra-class clustering.
        \vspace{-3mm}
        \begin{figure}[ht!]
          \centering
          \begin{subfigure}[b]{0.3\columnwidth}
            \includegraphics[width=\textwidth]{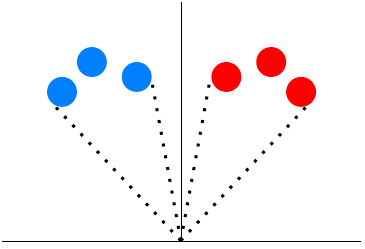}
          \end{subfigure}
          \hspace{3mm}
          \begin{subfigure}[b]{0.3\columnwidth}
            \includegraphics[width=\textwidth]{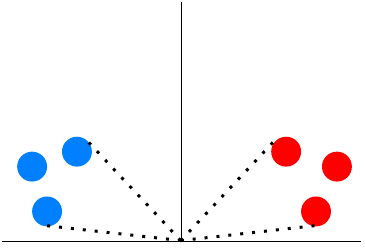}
          \end{subfigure}
          \vspace{-2mm}
          \caption{Low (left) and high (right) Inter-class separation}
          \label{fig:inter_sep}
        \end{figure}
        \vspace{-3mm}
    
    \item \textbf{Inter-class equidistance}: All embeddings from different classes are approximately equally spaced from each other. This allows removing hierarchies between classes, which is conceptually desirable since all errors have the same penalty in the classical classification problem.
        \vspace{-3mm}
        \begin{figure}[ht!]
          \centering
          \begin{subfigure}[b]{0.3\columnwidth}
            \includegraphics[width=\textwidth]{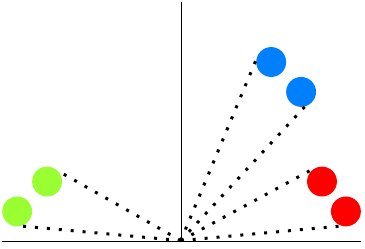}
          \end{subfigure}
          \hspace{3mm}
          \begin{subfigure}[b]{0.3\columnwidth}
            \includegraphics[width=\textwidth]{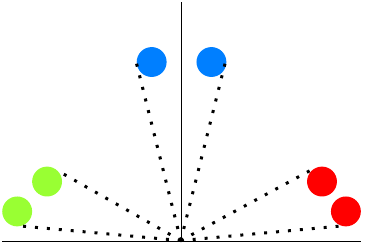}
          \end{subfigure}
          \vspace{-2mm}
          \caption{Low (left) and high (right) Inter-class equidistance}
          \label{fig:inter_equi}
        \end{figure}
        \vspace{-3mm}
\end{itemize}

Traditionally, only intra-class clustering and inter-class separation have been considered as desirable properties. However, we also consider it convenient to have intra-class and inter-class equidistance, since these allow to avoid intra-class and inter-class hierarchies, respectively, which is desirable in the canonical classification problem, since all errors and correct predictions are equally critical. As a result, as we will see in section \ref{sec:experiments}, maximizing these two improves the accuracy.

\subsection{Angular Distance Distribution Loss} \label{sec:add}
Having described the above properties and argued why they are desirable, we now present Angular Distance Distribution Loss, which encourages these properties. It imposes conditions on the first and second order statistical moments of the angular distances between embeddings. For now, we assume that the labels are hard, i.e. $y^{(i)}_k=1$ for one $k$ and 0 for the rest. With this idea, we can define the sets:
\begin{equation}
    D_p = \left\{ d_c\left( z^{(i)}, z^{(j)} \right)^2 \, \middle\lvert \,  y^{(i)} = y^{(j)} ; \, i\neq j \right\}
\end{equation}
\begin{equation}
    D_n = \left\{ \left( 1 - d_c\left( z^{(i)}, z^{(j)} \right) \right)^2  \, \middle\lvert \,  y^{(i)} \neq y^{(j)} \right\}
\end{equation}
where $d_c\left(x,y\right)=1-x^T \cdot y$, which takes values in the interval $[0,2]$, being $0$ when the two vectors are proportional, $1$ when they are orthogonal and $2$ when they are opposites. Next, we define the following terms from the previous ones:
\begin{equation}
    \mu_p = \frac{1}{|D_p|} \sum_{k\in D_p} k
\end{equation}
\begin{equation}
    \sigma_p = \sqrt{ \frac{ \sum_{k\in D_p} (k-\mu_p)^2 }{|D_p|-1} }
\end{equation}
and $\mu_n$ and $\sigma_n$ analogously for $D_n$. 
Each term can be related to one of the properties in \ref{subsec:properties} as follows:
\begin{itemize}
    \item Minimizing $\mu_p$ implies boosting intra-class clustering, since it implies minimizing the average distance between embeddings of the same class.
    \item Minimizing $\sigma_p$ implies promoting the intra-class equidistance, since we are reducing the variation of all the distances between embeddings of the same class.
    \item Minimizing $\mu_n$ implies boosting the inter-class separation, since we are promoting the embeddings of different classes to be orthogonal
    \item Minimizing $\sigma_n$ implies favoring the inter-class equidistance, since we are reducing the variation between embedding distances of different classes.
\end{itemize}
With all this, we define our loss function to minimize ADD as:
\begin{equation}
    L_{ADD} = \lambda_\mu^p \mu_p + \lambda_\sigma^p \sigma_p + \lambda_\mu^n \mu_n + \lambda_\sigma^n \sigma_n
\end{equation}
where $\bm{\lambda} = \{\lambda_\mu^p, \lambda_\sigma^p, \lambda_\mu^n, \lambda_\mu^n\}$ are hyperparameters. 
In section \ref{sec:experiments} we explore how each of these terms separately affects the accuracy and distribution of embeddings in space.

Finally, the loss function of our system is as follows:
\begin{equation}
    \mathcal{L} = \mathcal{L}_{CE} + \mathcal{L}_{ADD}
\end{equation}

\subsection{Soft Labels Adaptation}
In some scenarios, the labels used to optimize our model are soft, i.e., they represent the probability that an element belongs to each class instead of considering that an element belongs to a single class \cite{Martinmorato2023b}. One of the main causes of having soft labels is the use of data augmentation techniques such as mixup \cite{zhang2017mixup}. As mixup is widely used in audio classification \cite{zhang2018deep}, we propose a modification of our loss function to deal with soft labels.

When we have soft labels, it is still important to maximize intra-class clustering and intra-class equidistance, since we are interested that elements belonging to the same class should be close and at a similar distance from each other. However, inter-class separation must be reinterpreted, so that it would be desirable that if $y^{(i)}$ is more similar to $y^{(j)}$ than to $y^{(k)}$, then $z^{(i)}$ should be closer to $z^{(j)}$ than to $z^{(k)}$, and vice versa. For this, we must strive that $d_c\left(z^{(i)},z^{(j)}\right) = d_c\left(y^{(i)},y^{(j)}\right)$ for each pair $i,j$. To modify the loss function, we first define:
\begin{equation}
    \mathcal{L}_{\mu} = \frac{1}{N_B} \sum_{i\in B} \sum_{j\neq i} \left(  d_c\left(y^{(i)},y^{(j)}\right) - d_c\left(z^{(i)},z^{(j)}\right) \right)^2
\end{equation}
where $N_B = |B|(|B|-1)$ is the number of pairs in a batch. Optimizing $\mathcal{L}_{\mu}$ we manage to jointly maximize intra-class clustering and inter-class separation. In fact, we note that we do not lose generality with respect to the hard scenario, since $d_c\left(y^{(i)},y^{(j)}\right)$ holds 0 if $y^{(i)}=y^{(j)}$ and 1 otherwise and, therefore, optimizing $\mathcal{L}{\mu}$ is equivalent to optimizing $\lambda_\mu^p \mu_p + \lambda_\mu^n \mu_n$ with $\lambda_\mu^n = \frac{|D_n|}{|D_p|}\lambda_\mu^p$. 
In addition, since elements do not belong to a single class, it does not make sense to maximize the inter-class equidistance. Thus, when we have soft labels, we define the ADD as:
\begin{equation}
    L_{ADD}^{soft} = \lambda_\mu \mathcal{L}_{\mu} + \lambda_\sigma^p \sigma_p
\end{equation}

\vspace{-5pt}
\section{Experiments} \label{sec:experiments}

\subsection{Datasets}
\noindent\textbf{Environmental Sound Classification (ESC-50)} \cite{piczak2015esc} contains 2,000 5-second ambient sound recordings annotated with 5 classes, so that each audio belongs to a single class. In our experiments we follow the standard 5-fold cross-validation to evaluate our systems.

\noindent\textbf{Speech Commands V2 (KS2)} \cite{warden2018speech} is composed of 105,829 clips of 1-second spoken keywords annotated with 35 word classes. It is officially divided into 84,843, 9,981 and 11,005 clips for training, test and validation, respectively.

\noindent\textbf{IEMOCAP (ER)} \cite{busso2008iemocap} contains about 12 hours of speech with four different emotions. We use the standard 5-fold cross-validation proposed in \cite{yang2021superb} for evaluation.

\begin{table}[t]
\vspace{-3mm}
\centering
\caption{Hyperparam. per embeddings extractor and dataset}
\label{tab:hyperparameters}
\begin{tabular}{l|ccc|ccc}
\hline
\multicolumn{1}{c|}{} & \multicolumn{3}{c|}{AST}                            & \multicolumn{3}{c}{BEATs}                           \\
\multicolumn{1}{c|}{} & ESC              & KS2           & ER               & ESC           & KS2             & ER                \\ \hline
Window type           & \multicolumn{3}{c|}{Hanning}                        & \multicolumn{3}{c}{Povey}                           \\
Freq. Mask            & 24                  & 48            & 24            & \multicolumn{3}{c}{0}                              \\
Time Mask             & 96                  & 48            & 96            & \multicolumn{3}{c}{0}                              \\
Mixup $\lambda$       & 0                   & 0.5           & 0             & 0             & 0.5             & 0                \\
Epochs                & \multicolumn{3}{c|}{25}                             & \multicolumn{3}{c}{30}                             \\
Batch Size            & \multicolumn{3}{c|}{32}                             & \multicolumn{3}{c}{16}                             \\
Optimizer             & \multicolumn{3}{c|}{AdamW}                          & \multicolumn{3}{c}{Adam}                           \\
Learning rate         & 7e-4                & 6e-5          & 7e-4          & 8e-6          & 1e-4            & 8e-6              \\
 Momentum             & \multicolumn{3}{c|}{$\bm{\beta}=\{0.9, 0.98\}$}     & \multicolumn{3}{c}{$\bm{\beta}=\{0.95, 0.999\}$}   \\
Weight Decay          & \multicolumn{3}{c|}{1e-2}                           & \multicolumn{3}{c}{5e-6}                            \\ \hline
\end{tabular}
\end{table}

\begin{table*}[t]
\vspace{-1mm}
\centering
\caption{Accuracy for the different Datasets, Embeddings Extractors and Loss Functions}
\label{tab:acc}
\begin{tabular}{l|cc|cc|cc}
\hline
\multicolumn{1}{c|}{}               & \multicolumn{2}{c|}{ESC-50}                           & \multicolumn{2}{c|}{KS2}                      & \multicolumn{2}{c}{ER} \\
\multicolumn{1}{c|}{}               & AST                       & BEATs                     & AST                       & BEATs                     & AST                       & BEATs  \\ \hline
Cross-entropy                       & $93.97\pm 0.21$           & $91.05 \pm 0.41$          & $92.05 \pm 0.04$          & $88.94 \pm 0.13$          & $59.91 \pm 0.60$          & $61.66 \pm 0.31$          \\
Focal Loss \cite{lin2017focal}      & $94.40\pm 0.36$           & $91.10 \pm 0.49$          & -                         & -                         & $60.79 \pm 0.16$          & $62.17 \pm 0.05$          \\
OPL \cite{ranasinghe2021orthogonal} & $94.11\pm 0.37$           & $91.50 \pm 0.20$          & -                         & -                         & $60.53 \pm 0.42$          & $\mathbf{63.06 \pm 0.32}$  \\ \hline
ADD ($\bm{\lambda}=\{1,1,1,1\})$    & $\mathbf{94.68\pm0.09}$     & $\mathbf{92.22 \pm 0.06}$   & $\mathbf{97.54 \pm 0.06}$   & $\mathbf{90.49 \pm 0.16}$   & $\mathbf{61.30 \pm 0.38}$   & $62.73 \pm 0.17$          \\ \hline
\end{tabular}
\end{table*}

\begin{figure*}[ht!]
    \centering
    \captionsetup[subfigure]{justification=centering}
    \begin{subfigure}[b]{0.14\textwidth}
        \includegraphics[width=\textwidth]{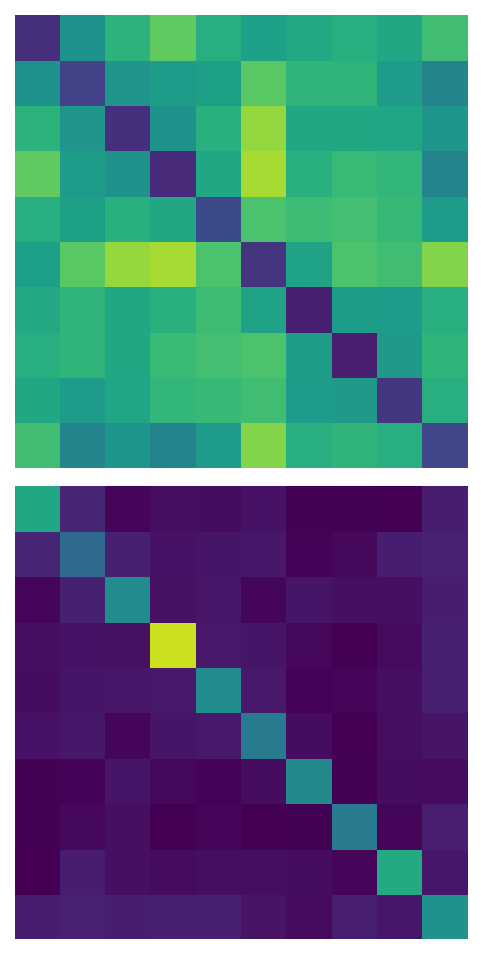}
        \caption{$\bm{\lambda}=\{1,0,0,0\}$ \\ $Acc=94.35\pm 0.23$}
        \label{subfig:cos_dist_1000}
    \end{subfigure}
    \hspace{0.005\textwidth}
    \begin{subfigure}[b]{0.14\textwidth}
        \includegraphics[width=\textwidth]{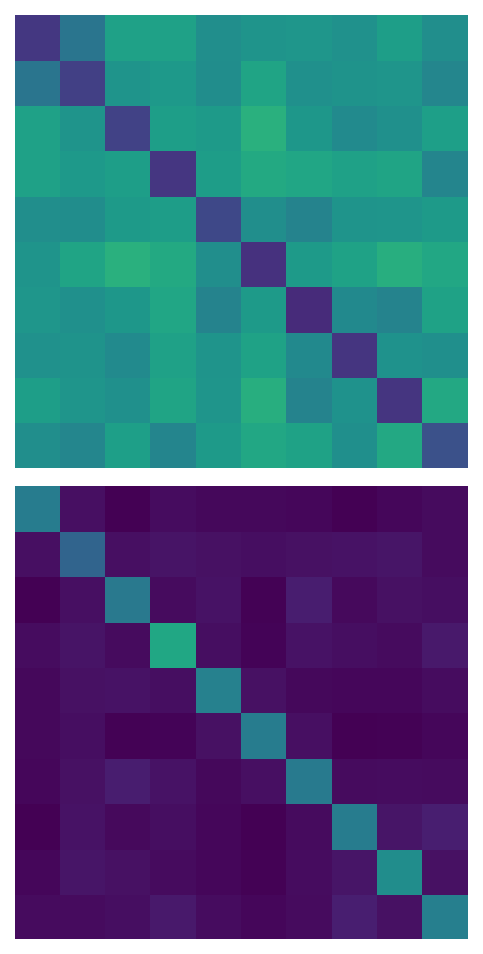}
        \caption{$\bm{\lambda}=\{0,1,0,0\}$ \\ $Acc=94.58\pm 0.38$}
        \label{subfig:cos_dist_0100}
    \end{subfigure}
    \hspace{0.005\textwidth}
    \begin{subfigure}[b]{0.14\textwidth}
        \includegraphics[width=\textwidth]{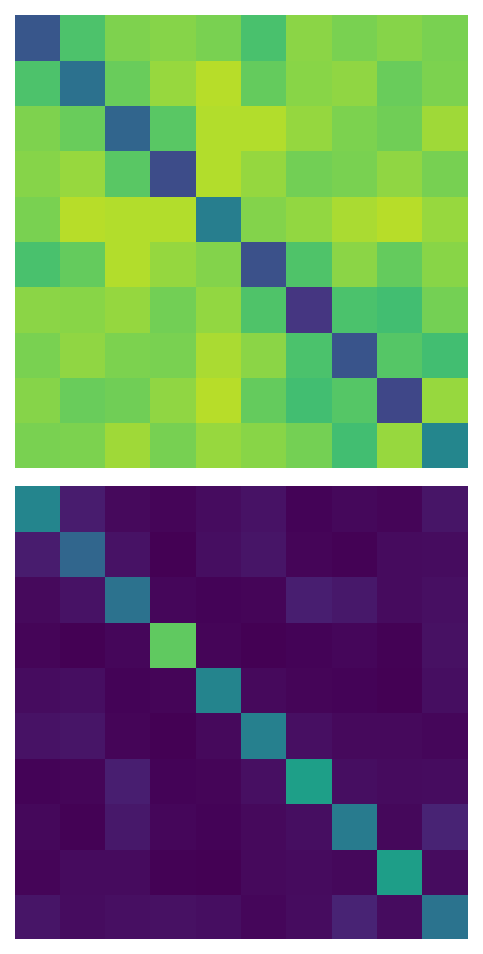}
        \caption{$\bm{\lambda}=\{0,0,1,0\}$ \\ $Acc=94.32\pm 0.13$}
        \label{subfig:cos_dist_0010}
    \end{subfigure}
    \hspace{0.005\textwidth}
    \begin{subfigure}[b]{0.14\textwidth}
        \includegraphics[width=\textwidth]{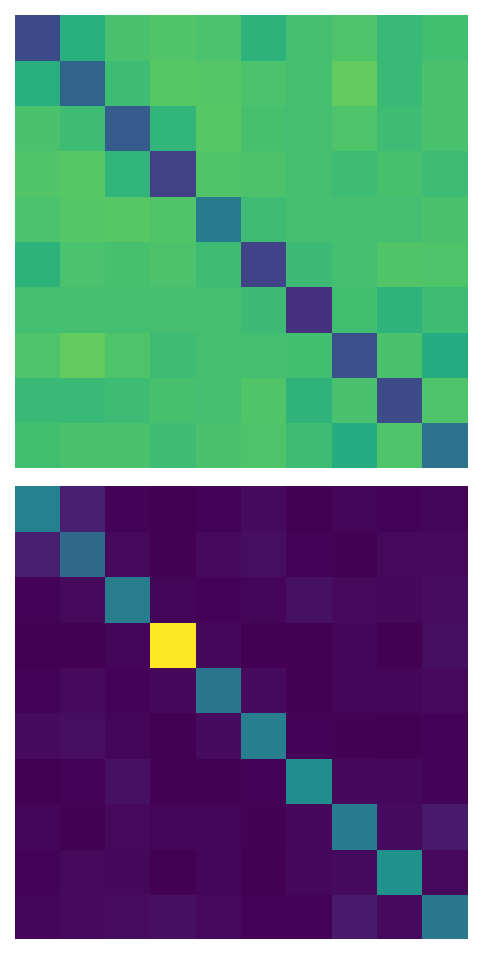}
        \caption{$\bm{\lambda}=\{0,0,0,1\}$ \\ $Acc=94.42\pm 0.18$}
        \label{subfig:cos_dist_0001}
    \end{subfigure}
    \hspace{0.005\textwidth}
    \begin{subfigure}[b]{0.14\textwidth}
        \includegraphics[width=\textwidth]{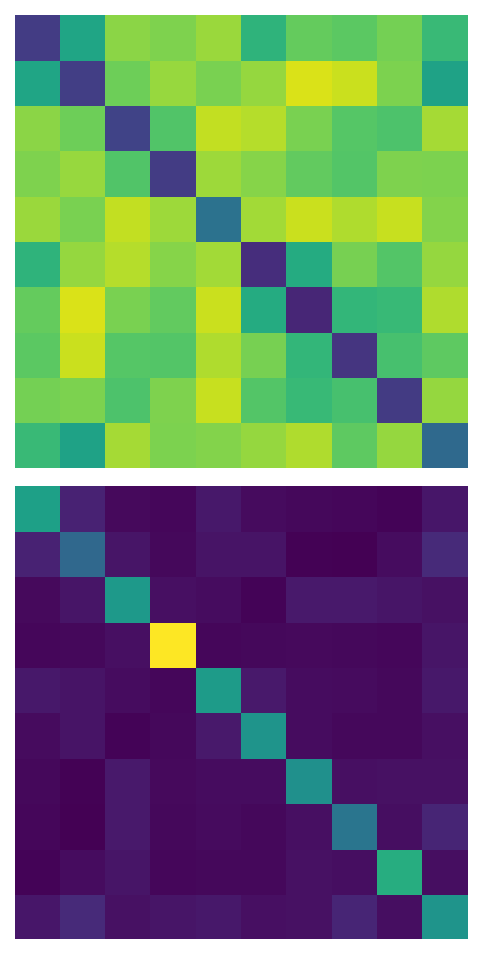}
        \caption{$\bm{\lambda}=\{1,0,1,0\}$ \\ $Acc=94.17\pm 0.26$}
        \label{subfig:cos_dist_1010}
    \end{subfigure}
    \hspace{0.005\textwidth}
    \begin{subfigure}[b]{0.14\textwidth}
        \includegraphics[width=\textwidth]{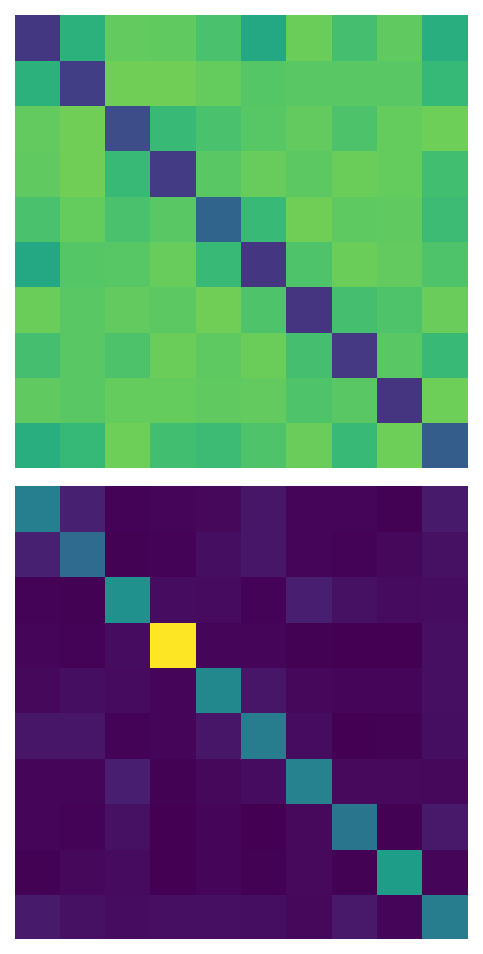}
        \caption{$\bm{\lambda}=\{1,1,1,1\}$ \\ $Acc=94.66\pm 0.37$}
        \label{subfig:cos_dist_1111}
    \end{subfigure}
    \begin{subfigure}[b]{0.5\textwidth}
        \includegraphics[width=\textwidth]{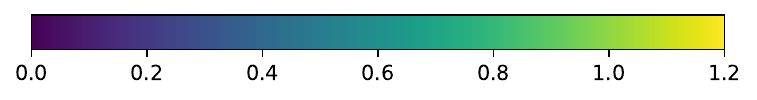}
    \end{subfigure}
    \vspace{-4mm}
    \caption{Mean (top row) and coefficient of variation (bottom row) of the $d_c$ values between embeddings of 10 classes of ESC-50. The accuracy given is calculated for the 50 classes. Coefficient of variation is defined as $\frac{\sigma}{\mu}$ and is used here instead of $\sigma$ because it normalizes the variation by normalizing by the mean, which changes depending on $\bm{\lambda}$, so it represents better intra-class equidistance.}
    \label{fig:cos_dist}
\end{figure*}

\subsection{Embeddings Extractors Architectures}
\noindent\textbf{Audio Spectrogram Transformer (AST)} \cite{gong2021ast} is the first to use Transformer type architectures for audio. The original AST model is pre-trained on Imagenet \cite{deng2009imagenet} and Audioset \cite{gemmeke2017audio}. We fine-tune it for each scenario.

\noindent\textbf{Bidirectional Encoder representation from Audio Transformers (BEATs)} \cite{chen2022beats} is a pre-training framework for learning representations from Audio Transformers, in which an acoustic tokenizer and a self-supervised audio model are optimized. We use the original pre-trained model with Audioset and finetune for each scenario.

\subsection{Hyperparameters}
For all our systems we use the audio signals at 16kHz. The input to the systems are 128 mel-spectrograms coefficients computed on 25 ms windows every 10 ms. We normalize the mean and standard deviation to 0 and 0.5, respectively. Some hyperparameters vary between scenarios. These details can be found in table \ref{tab:hyperparameters} and are inspired by the experiments in the original papers, with slight modifications due to computational limitations.

\subsection{Ablation Study on each term of the ADD}
We are going to analyze the influence of each of the terms of $\mathcal{L}_{ADD}$. First, we want to see if the hypotheses outlined in section \ref{sec:add} about the relationship between each ADD term and the properties of \ref{subsec:properties} hold. Second, we want to analyse the impact of each particular term in the ADD and their combination on the accuracy. For this, we train four classifiers, each with one of the elements of $\bm{\lambda}$ equal to 1 and the rest equal to zero. Third, we want to test whether optimizing intra-class and inter-class equidistance provides an advantage despite already optimizing intra-class clustering and inter-class separation. To do so, we train two classifiers: one with $\bm{\lambda}=\{1,0,1,0\}$ and another with $\bm{\lambda}=\{1,1,1,1\}$ and compare them. The dataset to be classified is ESC-50 and the embedding extractor used an AST in all cases. In figure \ref{fig:cos_dist} we present the mean and coefficient of variation of the $d_c$ between the embeddings of 10 pairs of classes and the accuracy calculated for all the classes.
\begin{itemize}
    \item In figure \ref{subfig:cos_dist_1000} we see that the distance between embeddings of the same class is in general the minimum in mean, i.e. the intra-class clustering is the maximum.
    \item In figure \ref{subfig:cos_dist_0100} we observe that the distances between the embeddings of the same class is the least spread, which means that the intra-class equidistance is the highest.
    \item In figure \ref{subfig:cos_dist_0010} we contemplate that the inter-class distance or separation is the maximum in mean.
    \item In figure \ref{subfig:cos_dist_0001} we find that the distances between embeddings of different classes are similar regardless of the class pairs, thus achieving a higher inter-class equidistance.
    \item In figure \ref{subfig:cos_dist_1010} we see that if we only optimize intra-class clustering and inter-class separation, the distances between different pairs of embeddings of different classes are very far from each other. In addition, there are classes for which embeddings are closer to each other than for other classes.
    \item In figure \ref{subfig:cos_dist_1111} we obtain embeddings that do not satisfy each property as well as when we try to optimize them separately, but with a good balance between all of them.
\end{itemize}
Finally, the best accuracy obtained is for $\bm{\lambda}=\{1,1,1,1\}$, that is, when we optimize all four properties together. This leads us to believe that all these properties have an influence in achieving a higher accuracy. 
In addition, we see that the properties that separately have most positively influence accuracy are inter-class and intra-class equidistance.

\vspace{-5pt}
\subsection{Quantitative Results}
We have found in the previous section that our loss function allows us to meet the properties that we consider desirable. Moreover, we have verified that for the analyzed scenario, the accuracy when the four properties are optimized jointly is better than when they are optimized separately. Here we perform a more extensive study in which we compare in terms of accuracy the ADD with other loss functions with good performance. We do not use Focal Loss and OPL for KS2, as these do not support soft labels and we use mixup for this dataset. We have performed all the experiments three times and we provide the mean and standard deviation of the accuracy. In table \ref{tab:acc} we can see that the results of our loss function is superior to the rest except in one case. This suggests that, in general, the described properties are desirable to improve accuracy and that the ADD function performs superiorly in different scenarios. 

\vspace{-5pt}
\section{Conclusions and Future Work}
In this paper we have presented four properties for embeddings of a classifier arguing why we consider these properties to be desirable. In addition, we have designed Angular Distance Distribution Loss, a loss function that is intended to allow us to obtain each of these properties. First, we have verified that, indeed, our loss function allows us to obtain emebddings that satisfy these properties separately and jointly. Subsequently, we have observed, for a given scenario, that the performance in terms of accuracy is better when all four properties are satisfied jointly than separately. Finally, we have found for different datasets and architectures that the fact that our embeddings satisfy these properties translates into better accuracy than other relevant loss functions in the literature. This validates the hypothesis about the importance of these properties to improve accuracy. 
We believe that the properties described in this work may be desirable also for other applications, such as anomaly detection or biometric recognition. Thus, experiments testing the ADD in these fields can be developed in the future

\bibliographystyle{IEEEtran}
\bibliography{refs}

%
%
%
%
%
%
%
%
%

\end{sloppy}
\end{document}